\documentclass[twocolumn,pre,superscriptaddress,preprintnumbers,amsmath,amssymb,longbibliography]{revtex4-1}

\usepackage{amsmath}    
\usepackage{graphicx}   
\usepackage{verbatim}   
\usepackage{color}      
\usepackage{subfigure}  
\usepackage{hyperref}   
\usepackage{dcolumn}
\usepackage{bm}
\usepackage[outdir=./]{epstopdf}
\usepackage{comment}
\usepackage[babel,autostyle=true,]{csquotes}

\newcommand{\mb}{\mathbf}

\begin{document}

\title{Dynamics of a droplet driven by an internal active device}
\author{R. Kree}
\author{L. R\"uckert}
\author{A. Zippelius}
\email{kree@theorie.physik.uni-goettingen.de}
\affiliation{ Institut f. Theoretische Physik, Universit\"at G\"ottingen, Friedrich-Hund Pl. 1, 37077 G\"ottingen, Germany}

\date{\today}

\begin{abstract}

  A liquid droplet, immersed into a Newtonian fluid, can be propelled
  solely by internal flow. In a simple model, this flow is generated
  by a collection of point forces, which represent externally actuated
  devices or model autonomous swimmers.  We work out the general
  framework to compute the self-propulsion of the droplet as a
  function of the actuating forces and their positions within the
  droplet.  A single point force, ${\mb F}$ with general orientation
  and position, ${\mb r}_0$, gives rise to both, translational and
  rotational motion of the droplet. We show that the translational
  mobility is anisotropic and the rotational mobility can be
  nonmonotonic as a function of $|{\mb r}_0|$, depending on the
  viscosity contrast. Due to the linearity of the Stokes equation,
  superposition can be used to discuss more complex arrays of point
  forces. We analyse force dipoles, such as a stresslet, a simple
  model of a biflagellate swimmer and a rotlet, representing a helical
  swimmer, driven by an external magnetic field. For a general force
  distribution with arbitrary high multipole moments the propulsion
  properties of the droplet depend only on a few low order multipoles:
  up to the quadrupole for translational and up to a special octopole
  for rotational motion. The coupled motion of droplet and device is
  discussed for a few exemplary cases. We show in particular that a
  biflagellate swimmer, modeled as a stresslet, achieves a steady
  comoving state, where the position of the device relative to the
  droplet remains fixed. In fact there are two fixpoints, symmetric
  with respect to the center of the droplet. A tiny external force
  selects one of them and allows to switch between forward and backward
  motion.

\end{abstract}

\pacs{47. 63. Gd, 87. 17. Jj, 87. 85. Tu}

\maketitle

\section{Introduction}
Micro- and nanoscale medical robotics is a rapidly emerging area of
research, which may open the way to many new and fascinating
applications like precision surgery, directed drug delivery,
micro-diagnostic sensing, uptake of toxins and many others (for recent
reviews see \cite{Li2017} and \cite{Hu2018}). A most important
challenge on the way towards reliable bio-technological systems is to
find bio-compatible, long lasting, and precisely controllable methods
of propulsion \textit{in vivo}.  Magnetically actuated helical
micro-motors on the $10 \mu m$ lengthscale, which are driven and
controlled by external fields provide a promising example.  They have
been used for important manipulations of soft materials, in particular
for steerable locomotion in small droplets, for the actuation of Human
B lymphocytes and the assembly or disassembly of complexes of droplets
and cells \cite{Ding2016}.  Furthermore they have already been
actuated in the peritoneal cavity of a mouse \cite{Servant2015} for
deep tissue analysis. Another promising technique is the biohybrid
actuation,
which uses molecular motors of biological systems as propulsion mechanism. The big advantage of this approach  is that  fuel is provided by the surrounding biofluid, and  these motors are highly optimised to convert the chemical energy into 3d propulsion\cite{Feinberg2015, Kojima2013}.

In many aspects of biohybrid systems can we profit from copying
evolutionary optimized designs from natural biological systems.  In
the case of self-propulsion \cite{Bechinger2016a}, the design of
artificial swimmers is inspired by biological microswimmers, such as
algae, bacteria and eukaryotic cells. Many of these have special
organelles, e.g. flagellae or cilia to propel the
microorganism. However, swimming without specialised organelles is
also found in nature. 
In the present work, we want to
explore possibilities of actuating a soft droplet by small internal
motors, which are either externally driven or operate autonomously.

Motion of a passive particle in the presence of a rigid cavity was first
considered by Oseen, who solved Stokes equation for the flow field of
a point force outside a sperical cavity~\cite{Oseen1927}. Several
extensions of Oseen's work have been obtained and other geometries,
such as plane walls and cylinders have been considered. For a more
recent exposition, see~\cite{Happel,KimKarrila2005}.  A Stokeslet near
a spherical viscous drop was first considered by Fuentes et
al.~\cite{Fuentes1988,Fuentes1989}. They solved the mobility problem
for both, the axisymmetric case~\cite{Fuentes1988}, as well as motion
perpendicular to the line connecting the
centers\cite{Fuentes1989}. More recently the oscillatory motion of a
particle inside an elastic cavity was
discussed~\cite{Loewen2018,Loewen2019}. Such a cavity is thought to
model a vesicle which is enclosed by a membrane with shear and bending
resistivity.  Whereas the early works focussed on colloidal
suspensions~\cite{Brady2011,Zia2016}, more recently active devices,
such as squirmers inside a droplet acquired attention.  In
\cite{Reigh2017}, the locomotion of a spherical squirmer encapsulated
inside a droplet of comparable size suspended in another viscous fluid
has been studied.  The authors show that the encaged swimmer is able
to propel the droplet, and in some situations both remain in a stable
co-swimming state. In~\cite{Shaik2018}, the coswimming of a squirmer
in a droplet with a nonuniform surface tension was studied. The latter
provides an additional mechanism for self-propulsion and can increase
or decrease both velocities, that of the squirmer and that of the
droplet. Both papers~\cite{Reigh2017,Shaik2018} consider only
axisymmetric configurations, resulting in translations.

Here we consider a spherical droplet which is immersed into an ambient
Newtonian fluid and actuated by small internal motors. Our focus
is the propulsion of the droplet due to an internal active device. We
assume the device to be small as compared to the droplet and hence
model it by a collection of point forces~\cite{Felderhof2012}. In contrast to ~\cite{Reigh2017,Shaik2018}, we consider asymmetric configurations and compute both, 
the linear and rotational velocity of the droplet generated by point
forces. Quantitative results are obtained for both cases, an
autonomous, force and torque free swimmer and an externally actuated
device.  In the next section (\ref{model}) we introduce the model and
obtain its analytical solution in sec.~\ref{analytics}. We present
results for a single point force, representing an externally driven
device, as well as for a dipolar and a quadupolar force configuration,
representing autonomous swimmers in sec.~\ref{results}. The coupled dynamics of droplet and device is discussed for exemplary cases in
sec.~\ref{sec:mobility}. Conclusions are presented in
sec.~\ref{conclusions}; details of the analytical calculations are
defered to the appendices.
\section{Model}
\label{model}
We want to study the propulsion of a droplet, which is driven by a
device, which is either contolled externally or may be autonomous. The
droplet is assumed to be spherical and consists of an incompressible
Newtonian fluid with viscosity $\eta^-$. It is immersed into an
ambient Newtonian fluid of viscosity $\eta^+$ which is at rest in the
laboratory frame (LF). The two fluids are assumed to be completely
immiscible with density $\rho=1$, so that the droplet is neutrally
bouyant.

For small Reynolds number the flow field created by the moving device can be calculated from Stokes's equation 
\begin{equation}
\label{eq:stokes}
\nabla\cdot\boldsymbol{\sigma}=\eta\nabla^2\mb{v}-\nabla p=-\mb{f} , 
\end{equation}
supplemented by the incompressibility condition $\nabla\cdot\mb{v}=0$.
The viscosity $\eta$ in Eq. (\ref{eq:stokes})  jumps between $\eta^+$ and $\eta^-$ across the boundary of the droplet.
The viscous stress tensor $\boldsymbol{\sigma}$ is given by its
cartesian components
$\sigma_{ij}=-p\delta_{ij}+\eta(\partial_iv_j+\partial_jv_i)$, with
the pressure $p$ determined from the incompressibility.
The force density exerted by the active device is denoted by $\mb{f}$ and will be specified below.
On the
boundary of the droplet we assume continuity of the flow field
$\mb{v}(\mb{r})$ and of the tangential stress, whereas the normal
stress jumps due to a homogeneous surface tension $\gamma_0$, so that
$\mb{e}_r\cdot(\boldsymbol{\sigma}_+-\boldsymbol{\sigma}_-)=2\gamma_0\mb{e}_r$.
Once the internal flow $\mb{v}$ has been computed, the linear and angular momentum of the droplet follow from
\begin{equation}
  \label{centerofmass}
  M\mb{v}_{CM}=\int_V d^3x \;\mb{v}, \quad
  I\boldsymbol{\omega}=\int_Vd^3x \;\mb{r}\times\mb{v}.
\end{equation}
with total mass $M=4\pi/3$ and moment of inertia $I=8\pi/15$. 
Here the
integral is over the volume of the droplet, $V$, and here and in the following
we use the droplet`s radius as the unit of length.

We model the simplest externally controlled device by a point force:
\begin{equation}
  \mb{f}(\mb{r})=\mb{F}\delta(\mb{r}-\mb{r}_0).
\end{equation}
Once the solution for the point force has been constructed, more
general force distributions can be treated by superposition of the
flow fields, because the Stokes equations are linear in $\mb v$. Of
particular interest are force dipoles and force quadrupoles which can
serve as models for an autonomous force free swimmer. Another
motivation for considering a point force stems from the following
well-known fact~\cite{KimKarrila2005}: The flow field of a moving
sphere of radius $a$ is correctly represented by a point force and a
point quadrupole in unbounded space. Even though this result is not
expected to hold for a sphere within a finite droplet, it may serve as
an approximation provided the particle radius is small compared to
all other length scales~\cite{Felderhof2012}. This point will be discussed in more detail in sec.\ref{sec:mobility}.

\section{Analytical Solution }
\label{analytics}
Our general strategy is to construct a special solution of the
inhomogeneous equation and then add a homogeneous solution to match
the boundary conditions.  As a special solution of the inhomogeneous
problem we can choose the classical Oseen tensor solution of a point
force at position $\mb{r}_0$ in an unbounded fluid~\cite{KimKarrila2005}:
\begin{equation}
8\pi\eta\mb{G}_{ij}(\mb{r}-\mb{r}_0)=\frac{1}{|\mb{r}-\mb{r}_0|}\delta_{ij}+\frac{(\mb{r}-\mb{r}_0)_i(\mb{r}-\mb{r}_0)_j}{|\mb{r}-\mb{r}_0|^3}\nonumber.
\end{equation} 
However, the usual
representation of this solution is not easy to match to boundary
conditions on the surface of a sphere. Instead of expanding the Oseen
tensor 
into solutions of the homogeneous Stokes equation, we prefer to construct the
solution from Stokes equation in terms of vector spherical harmonics (VSH)
directly.
Our choice of VSH is $\mb{Y}_{lm}^{(0)}=\mb{e}_rY_{lm}, \,\mb{Y}_{lm}^{(1)}=r\mb{\nabla} Y_{lm} $
and $\mb{Y}_{lm}^{(2)}=\mb{e}_r \times\mb{Y}_{lm}^{(1)}$.
The vector sperical harmonics form a complete orthogonal
set on the surface of a unit sphere with respect to the scalar product
 \begin{equation}
 (\mathbf{h}, \mathbf{g})= \int \text{d} \Omega ~ \mathbf{h}^*(\Omega) \cdot \mathbf{g}(\Omega).
 \label{eq:scalarproduct}
 \end{equation}    
 While the $\mb{Y}_{lm}^{(0)}$ have norm 1, the $s=1,2$ fields have a norm of $(A_l^{(s)})^{-1}=\ell (\ell+1)$. 
 For further properties of these functions we refer the reader to \cite{Barrera1985}.

The expansion of the force density is explicitly given by
\begin{equation}
  \label{force_expansion}
  \mb{F}\delta(\mb{r}-\mb{r}_0) =
                              \frac{\delta(r-r_0)}{r_0^2} \sum_{s=0}^2\sum_{lm} f_{lms}(\Omega_0)\mb{Y}_{lm}^{(s)}(\Omega) 
\end{equation}
with
$ f_{lms}(\Omega_0) = A_l^{(s)} [\mb{Y}_{lm}^{(s)}(\Omega_0)]^*\cdot
\mb{F}$.
The flow field and the pressure
which are generated by the above point force are similarly expanded in VSH
\begin{align}
  \label{flow_expansion}
 \mb{v}(\mb{r}|\mb{r}_0)&= \sum_{s=0}^2\sum_{lm} v_{lms}(r,\mb{r}_0)\mb{Y}_{lm}^{(s)}(\Omega) \\
  p(\mb{r}|\mb{r}_0)&=\sum\limits_{lm} p_{lm}(r,\mb{r}_0) Y_{lm}(\Omega).
                      \label{pressure_expansion}
\end{align}

The $\{v_{lms}\}$ and $\{p_{lm}\}$ are constructed
by a superposition of a special solution of the inhomogeneous equation
and the general solution of the homogeneous equation in order to
satisfy the boundary conditions.
This provides an exact analytic solution of the flow field of our model in terms of an infinite series. The applicability of this result may be limited by the rate of convergence of the series.
For our purposes of calculating the propulsion velocities of the droplet (Eq.\ref{centerofmass}), however,  it proves to be very convenient, because the total momentum and angular momentum of the droplet are determined  exclusively by the $l=1$ component.
Inserting the expansion of $\mb{v}$  into Eq.\ref{centerofmass} reveals (see Ref. \cite{Kree_2017}): 
\begin{align}
    \mb{v}_{CM} & =\frac{3}{4\pi}\sum_m\int_{\partial V}d\Omega \;v_{1m0}(r=1,\mb{r}_0)
                Y_{1m}(\Omega)\mb{e}_r\label{eq:vcm}\\
\boldsymbol{\omega} & =-\frac{15}{8\pi}\sum_m\int_{V}d^3r\;r^2\;v_{1m2}(r,\mb{r}_0)
                \mb{\nabla}Y_{1m}(\Omega).\label{eq:omega}          
\end{align}

The homogeneous solutions to the Stokes equation are well known and
recapitulated in  the appendix. Here we concentrate on the inhomogeneous equation and demonstrate our strategy to find a special solution for the pressure. The derivation of the inhomogeneous flow are delegated to the appendix.

Substituting the expansions of Eqs.~(\ref{force_expansion},\ref{pressure_expansion}) into $\nabla^2 p=\nabla\cdot\mb{f}$, yields
 \begin{align}
 \label{eq:pressure}
 &\frac{\text{d}^2 p_{lm}}{\text{d}r^2}+\frac{2}{r}\frac{\text{d}p_{lm}}{\text{d}r}-\frac{l(l+1)}{r^2}p_{lm}=
   \frac{f_{lm0}}{r_0^2} \frac{\text{d}}{\text{d}r}\delta(r-r_0)\nonumber\\
   &+\frac{2f_{lm0}}{r_0^3}\delta(r-r_0)-\frac{l(l+1)f_{lm1}}{r_0^3}\delta(r-r_0).
 \end{align}
 We construct a special solution of the inhomogeneous equation (Eq.\ref{eq:pressure}) with help of the inner and outer solutions with respect to the position of the point force ${\mb r}_0$. We thus use the ansatz:
 \begin{align}\label{p1}
 p_{lm}^{inh}(r)=A_{lm} r^{\ell} \Theta ( r_0 - r) + B_{lm} r^{-\ell-1} \Theta(r-r_0) ~.
 \end{align}
 and plug this ansatz into Eq.(\ref{eq:pressure}) to determine the values of
 $A_{lm}$ and $B_{lm}$,
\begin{align}\label{p2}
 A_{lm} &= \frac{- (\ell+1) f_{lm0} + \ell (\ell+1) f_{lm1}}{2\ell +1} r_0^{-\ell -2} \\
 B_{lm} &= \frac{\ell f_{lm0} + \ell (\ell+1) f_{lm1}}{2\ell +1} r_0^{\ell -1}.
\end{align}
Subsequently we add the general solution of the homogeneous equation,
$p_{lm}^{hom}(r)\propto r^l$, to fulfil the boundary conditions at the droplet's surface.

The flow fields are constructed in analogy as detailed in the appendix.
Here we just give the complete solution for the rotational velocity in order to illustrate the general type of flow field generated by a point force:  
\begin{align}
  \label{chiral_v}
  v_{1m2}(r,\mb{r}_0) & = \frac{f_{1m2}(\Omega_0)}{3\eta^-}\big(
                        \frac{\eta^+-\eta^-}{\eta^+}r r_0 \nonumber\\
                      & -\Theta(r-r_0)\frac{r_0}{r^2}                                                                -\Theta(r_0-r)\frac{r}{r_0^2} \big)\\  
 \label{chiral_f} 
  f_{1m2}(\Omega_0) & =\mb{F}\cdot (\mb{r_0}\times \mb{\nabla})
                      Y_{1m}(\Omega_0)/2
  \end{align}
and a similar expression for $v_{1m0}(r,\mb{r}_0)$.

\section{Results}
\label{results}

The general formalism of the previous section allows us to give simple
and exact expressions for the translational and rotational velocities
of a droplet, which is driven by a configuration of point forces. This
is the central finding of the present work. In this section we first
give basic results for the propulsion velocities $\mb{v}_{cm}$ and
$\boldsymbol{\omega}$ due to a single point force.  All other configurations
can be studied by simple linear superpositions of these velocities. As
examples we consider pairs of point forces, such as stresslets and
rotlets, and a device made of three linearly arranged points
(see Fig.1). The examples have been chosen to provide crude
approximations of biflagellate microorganisms and of helical magnetic
swimmers.  Finally we discuss arbitrary configurations and show that
only a few low order multipoles of the force distribution determine
the propulsion velocities completely.
\begin{figure}\label{Fig1}
\includegraphics[width=0.3\textwidth]{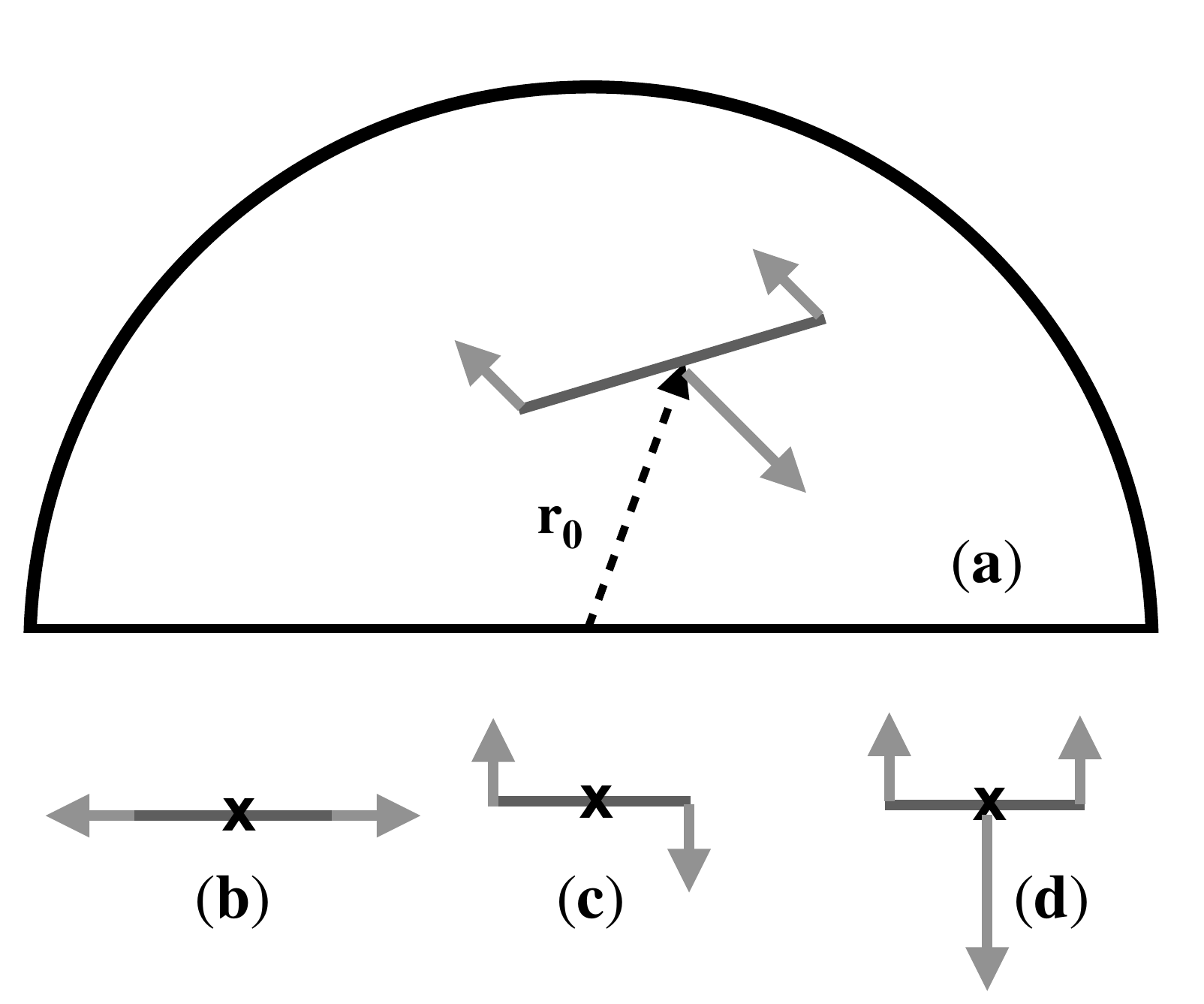}
\caption{a) Upper half of the droplet with examples of devices of point forces at $\mb{r}_0$; b) stresslet; c) rotlet; d) quadrupolar device; the point forces are indicated by vectors.}
\end{figure}

\subsection{Single point force}

First consider a single point force
$\mb{f}(\mb{r})= \mb{F}\delta(\mb{r}-\mb{r}_0)$, representing an
externally driven device.  Using Eqs.(C1,C2,C4-C7)
we find
\begin{equation}
\mb{v}_{cm}= \frac{1}{4\pi \eta^+} \frac{1}{3\lambda +2}\Big( (2\lambda +3) \mb{F} - r_0^2 (2\mb{F} -\mb{F}_{||})\Big) =\mb{\mu}_{t}(\mb{r}_0) \mb{F} .
\label{eq:translationV}
\end{equation}
Here we have introduced the vector component of the force parallel to
$\mb{r}_0$, $\mb{F}_{||}=(\mb{F}\cdot\mb{r}_0)\mb{r}_0/r_0^2$, and the
viscosity contrast $\lambda=\eta^-/\eta^+$. Note that the
translational mobility tensor $\mb{\mu}_{t}$ in
Eq.(\ref{eq:translationV}) is {\it anisotropic}.

From Eqs.(B4,B5,C8,C9) 
the angular velocity is obtained in the form
\begin{equation}
\boldsymbol{\omega} =- \frac{\mb{r}_0 \times \mb{F}}{16\pi \eta^-}\Big(2\lambda + 3(1-r_0^2)\Big)=-\mu_{rot}(r_0)\, \mb{r}_0 \times \mb{F}.
\label{eq:rotationOmega}
\end{equation}
The rotational mobility $\mu_{rot}$  in Eq.(\ref{eq:rotationOmega}) is isotropic and decreasing with $r_0$ while 
the torque  is increasing. This may lead to a non-monotonic dependence of
$\boldsymbol{\omega}(r_0)$. For example,  a point force $\mb{F}=F\mb{e}_x$ located on the z-axis at $\mb{r}_0=z_0\mb{e}_z$ causes an angular frequency
$\boldsymbol{\omega}=\omega({z}_0)\mb{e}_y$, which  takes on a maximum at $z^*=\sqrt{2\lambda +3}/3$, which is inside the droplet if $\lambda <3.$ 
Note that a single point force will in general lead to both translation and rotation of the droplet.  

Rotation is prevented, if the torque with respect to the center of the droplet vanishes, i.e. if the force is parallel to $\mb{r}_0$. 
Choosing $\mb{F}= F\mb{e}_z$ and $\mb{r}=z_0\mb{e}_z$ the droplet moves in z-direction with a mobility (see equation(C3)), which can be written in the form
\begin{equation}
\mu(z_0)=\mu_{HR} + \frac{1}{4\pi} \frac{1-z_0^2}{2+3\lambda},
\end{equation}
where
$\mu_{HR}= (1/2\pi\eta^+)((\lambda+1)/(2+3\lambda))$ is the classical result of Hadamard and Rybshinski for a spherical droplet moving with precribed velocity and driven by interface tractions \cite{KimKarrila2005}. Note that $\mu(z_0) \geq \mu_{HR}$ and it approaches $\mu_{HR}$ if $z_0\to 1$.

\subsection{Force dipoles}\label{sec:dipole}
Next consider  pairs of point forces
$\mb{f}^\pm(\mb{r})= \mb{F}^\pm\delta(\mb{r}-\mb{r}^\pm)$
with $\mb{F}^\pm=\pm \mb{F}$, located at $\mb{r}^\pm=\mb{r}_0\pm \mb{d}/2$,
which make up force dipoles 
$\mb{f}(\mb{r})=\mb{f}^{+}(\mb{r}) + \mb{f}^{-}(\mb{r})$. For such force dipoles the total force vanishes, but there is in general a nonzero torque.
By elementary vector algebra one finds from  Eqs.(\ref{eq:translationV}, \ref{eq:rotationOmega}) that
\begin{equation}
\mb{v}_{cm} = \frac{1}{4\pi\eta^+(3\lambda +2)}\Big((\mb{F}\cdot\mb{d})\mb{r}_0 + (\mb{F}\cdot\mb{r}_0)\mb{d}-4(\mb{r}_0\cdot\mb{d})\mb{F}\Big)
\label{eq:vcmPairs}
\end{equation}
and

\begin{align}
  \mb{\omega}&= \frac{1}{16\pi\eta^-}\left[3\left(r_0^2 + \frac{d^2}{4}\right) -2\lambda -3\right] (\mb{d}\times\mb{F})\nonumber\\
  &+ \frac{6}{16\pi\eta^-}\big(\mb{r}_0\cdot\mb{d} \big)\, (\mb{r}_0\times \mb{F}).
\label{eq:omegaPairs}
\end{align}
Point-like force dipoles are included in these results by taking the
limit $d \to 0$ while keeping the dipole strength $Fd$ fixed.

The simplest example of an autonomous device is a {\it stresslet}-like force pair characterized by
$\mb{F}= F \mb{d}/d$ (see Fig.1b). This may be considered as a crude approximation  of a biflagellate microorganism~\cite{Mehanda2008} 
with a thrust $\mb{F}$ exerted by flagella and balanced by the viscous drag
$-\mb{F}$ of the cell body. Note that the 
symmetric configuration studied here would not lead to self-propulsion of the device in free space.   

The droplet's translational velocity becomes 
\begin{equation}
\mb{v}_{cm}=  \frac{Fd}{4\pi\eta^+(3\lambda +2)}\big(1-3\mb{P}_{||}(\mb{d}) \big)\,\mb{r}_0.
\end{equation} 
Here, $\mb{P}_{||}(\mb{d})$ denotes the projector in the direction of $\mb{d}$.
Although the stresslet's intrinsic torque is zero, it will  generate rotational motion of the droplet  with
\begin{equation}
\boldsymbol{\omega} =  \frac{3}{8\pi\eta^-} \big(\mb{r}_0\cdot\mb{d} \big)\, (\mb{r}_0\times \mb{F}).
\end{equation}
This angular velocity is independent of the exterior viscosity
$\eta^+$; it only vanishes if $\mb{d}$ is perpendicular to $\mb{r}_0$
or if $\mb{F}$ is parallel to $\mb{r}_0$.

A {\it rotlet}-like force pair $\mb{F} \perp \mb{d}$ (see Fig.1c) may
be considered as a first approximation to a helical magnetic swimmer
driven by a rotating field~\cite{Dreyfus2005},\cite{Ding2016} and
exerting a finite torque on the fluid. The propulsion velocities are
easily read off from Eqs.(\ref{eq:vcmPairs}, \ref{eq:omegaPairs}) and
show that the droplet is not only rotated but also translated by this
force pair.  If, for example, $\mb{d}\, ||\, \mb{e}_x$ and
$\mb{F}\, ||\, \mb{r}_0 = z_0 \mb{e}_z$ the droplet will move in
x-direction with speed $v_{cm}= Fdz_0/(4\pi\eta^+(3\lambda +2))$.

\subsection{Quadrupolar, autonomous device}\label{sec:quadrupole}
As a last example of simple point force configurations consider a
slightly refined model of a biflagellate microorganism characterised by
$\mb{f}(\mb{r})=2\mb{F}\delta(\mb{r}-\mb{r}_0) -
\mb{F}\delta(\mb{r}-\mb{r}_0 - \mb{d}) - \mb{F}\delta(\mb{r}-\mb{r}_0
+ \mb{d})$ (see Fig.1d). Here the two flagellae are described as separate,
symmetrically arranged point forces~\cite{Dolger2017}. The device is
fully autonomous, because both, the total force and the torque with
respect to the point $\mb{r}_0$,
vanish. 
The propulsion velocities become
\begin{equation}
\mb{v}_{cm}=  \frac{1}{4\pi \eta^+} \frac{1}{3\lambda +2}\Big(4d^2 \mb{F} -(\mb{d}\cdot\mb{F})\mb{d}\Big)
\end{equation}
and 
\begin{equation}
\boldsymbol{\omega} = \frac{3}{8\pi\eta^-} \Big((2\mb{r}_0\cdot\mb{d})(\mb{d}\times\mb{F})+ d^2(\mb{r}_0\times\mb{F})\Big)
\end{equation}
Note that unlike a single point force or a force pair the translation
velocity of the 3-point device is independent of the position
$\mb{r}_0$ of the configuration in the droplet's interior. The droplet
is translated without rotation if $\mb{F} \, || \, \mb{r}_0$ and
$\mb{d} \, \perp \, \mb{r}_0$ holds.

\subsection{Droplet driven by a general configuration of point forces}

It is straightforward to generalise the results of the previous
subsection to arbitrary configurations
$\mb{f}(\mb{r})=\sum_{\nu=1}^M
\mb{F}^{(\nu)}\delta(\mb{r}-\mb{r}^{(\nu)})$ with
$\mb{r}^{(\nu)}=\mb{r}_0 + \mb{d}^{(\nu)}$. The point $\mb{r}_0$,
which here appears as an arbitrary marking of the position of the
device, may acquire a physical meaning in more detailed models as will
be discussed in sec.(\ref{sec:mobility}). Superposition of the
contributions of point forces gives the propulsion velocities

\begin{equation}
\mb{v}_{cm}= \frac{1}{4\pi \eta^+} \frac{1}{3\lambda +2}\Big( (2\lambda +3) \mb{F}_M -\sum_{\nu=1}^M (r^{(\nu)})^2 \big(2\mb{F}^{(\nu)}-\mb{F}_{||}^{(\nu)}\big)\Big)
\label{eq:vcmGeneral}
\end{equation}
and
\begin{align}
  \boldsymbol{\omega} = -\frac{1}{16\pi \eta^-}\Big(&(2\lambda +3)(\mb{N}_M +
 \mb{r}_0\times\mb{F_M})\nonumber\\
  &- 3 \sum_{\nu} (r^{(\nu)})^2\mb{r}^{(\nu)} \times \mb{F}^{(\nu)}) \Big),
\label{eq:omegaGeneral}
\end{align}
which are determined by only a few low order multipoles. It is easily
seen by inserting $\mb{r}^{(\nu)}=\mb{r}_0 +\mb{d}^{(\nu)}$ into
Eqs. (\ref{eq:vcmGeneral}, {\ref{eq:omegaGeneral}}) that
$\mb{v}_{cm}$ 
is completely fixed by the total force
$\mb{F}_M=\sum_{\nu=1}^M \mb{F}^{(\nu)}$, the second rank tensor of
force dipole moments $\mb{D}=\sum_\nu \mb{d}^{(\nu)} \mb{F}^{(\nu)} $
(including the total torque
$\mb{N}_M=\sum_{\nu=1}^M \mb{d}^{(\nu)}\times\mb{F}^{(\nu)}$ ) and the
third rank tensor of quadrupole moments
$\mb{Q}=\sum_\nu \mb{d}^{(\nu)} \mb{d}^{(\nu)}\mb{ F^{(\nu)}}$.  For
the angular velocity, $\mb{\omega}$, a special octopole moment
$\mb{W}=\sum_\nu (\mb{d}^{(\nu)}\cdot\mb{d}^{(\nu)})\,\mb{d}^{(\nu)}
\mb{F}^{(\nu)}$ is needed in addition.  Therefore different
distributions of internal forces may generate the same propulsion
velocities.  In particular, if the configuration is autonomous and its
dipole moment vanishes, the translation velocity is independent of the
position $\mb{r}_0$ (the 3-point device of the previous subsection is
a simple example).  If in addition $\mb{Q}=\mb{0}$ but $\mb{W}$ is
finite then $\mb{v}_{cm}=\mb{0}$ and the droplet is only rotated. All
higher order multipole moments beyond the octopole do not generate any
propulsion of the droplet.

\section{Coupled dynamics of  device and droplet}
\label{sec:mobility}
So far we have discussed snapshots of droplet and device, the latter being treated as an assembly of point forces. The motion of a real 
device within the droplet depends upon the details of its material properties, its structure and its internal dynamics. A discussion of the general case is beyond the scope of this work. In the following we illustrate the use of our results to determine the coupled evolution of the droplet and devices consisting of small spherical beads. We consider a single bead dragged by an external force and a simple steering problem of a biflagellate model. 

\subsection{Small bead dragged by an external force}
A spherical particle of radius $a$  is dragged along the z-axis with an external force $\mb{F}^{ext}= F \mb{e}_z$. 
The current position of the center of the sphere is $\mb{r}_0=z_0\mb{e}_z$. In free space its velocity is given by Stokes law as $\mb{U}_{device}=\mb{F}^{ext}/ (6a\pi\eta^-)=\mu_0 \mb{F}^{ext}$.  In the droplet the moving sphere causes a response  flow $\mb{v}^{(a)}(\mb{r}|\mb{r}_0)$ due to the interface and the device velocity can be obtained from Fax\'{e}n's law in the explicit form
\begin{equation}
\mb{U}_{device}= \mu_0\mb{F}^{ext} + \left(1 + \frac{a^2}{6}\nabla^2\right)\mb{v}^{(a)}(\mb{r}_0|\mb{r}_0),
\label{eq:UdeviceSphere}
\end{equation}
provided the response flow, $\mb{v}^{(a)}(\mb{r}|\mb{r}_0)$, is known in the vicinity of $\mb{r}\sim\mb{r}_0$.   In the point-force framework considered here, we  cannot calculate  $\mb{v}^{(a)}$ exactly, but for a small sphere $(a \ll 1)$, with distance $D \gg a$ from the interface, we can approximate the flow it generates at the interface by that of its leading multipole, i.e. the point force
$\mb{F}^{ext}$.  This far field determines the resulting response flow, which  has been calculated in the Appendix~\ref{app:greensfct}, and which we use in equation (\ref{eq:UdeviceSphere}).  As the flow is smooth at $\mb{r}_0$,  we can drop the derivative term $\propto a^2 \nabla^2 $ to leading order in $a$.

For the axially symmetric structure of the device the calculations
simplify considerably. The response flow is parallel to the z-axis and
leads to a change of the particle mobility,
$\mb{U}_{device}= (\mu_0 + \delta\mu_0(z_0))F\mb{e}_z $ with
\begin{equation}
\delta\mu_0(z_0)F=  \sum_{l=1}^\infty \left(\frac{2l+1}{4\pi}\right)^{1/2} \left(a_{l0}^- z_0^{l+1} + b_{l0}^- z_0^{l-1}\right).
\label{eq:Uparticle}
\end{equation}
Here $a_{lm}^-, b_{lm}^-$ are defined in equations \ref{vlm0},\ref{vlm1}.
The device velocity relative to the droplet velocity, $\mb{U}_{device}-\mb{v}_{cm}$, is always positive. It is minimal for the device in the center of the droplet and grows as the device moves towards the interface. For an illustrative example with droplet radius $a=0.05$ and a viscosity contrast $\lambda=30$, the difference is a few percent.

If we want to integrate the device velocity to determine its trajectory we have to assume that the droplet retains its spherical shape. As the flow field on the interface contains shape-changing components, these have to be counteracted by a large surface tension, which prevents any changes in area, i.e we have to assume a sufficiently small capillary number \cite{Vlahovska2009}.

\subsection{Externally guided biflagellate}
\begin{figure}
\includegraphics[width=0.45\textwidth]{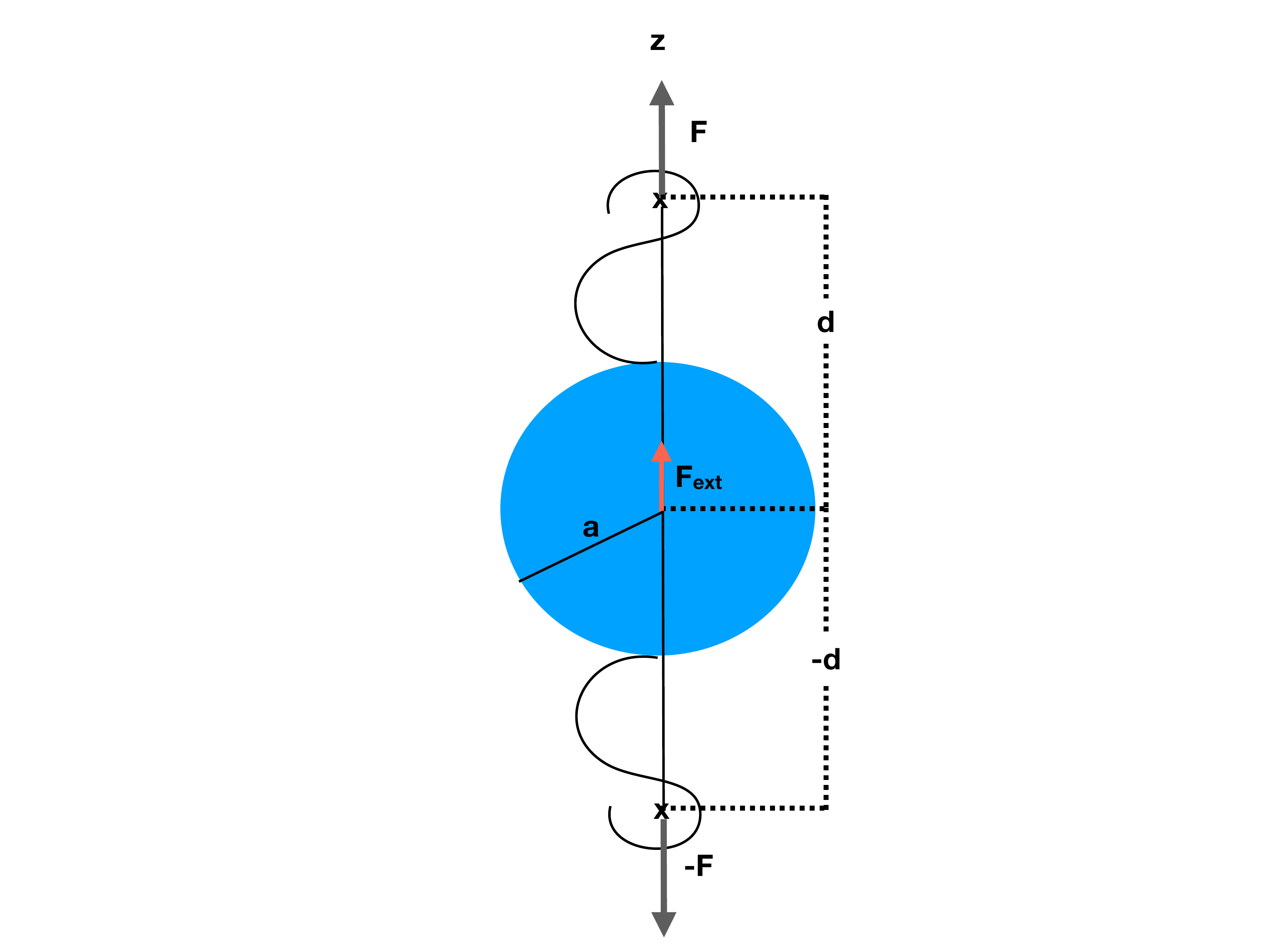}
\caption{\label{Fig2}Biflagellate swimmer, modeled by a small sphere with an attached stresslet and subject to an external force $\mb{F}^{ext}$}
\end{figure}

To drive a spherical droplet along a prescribed trajectory by an
internal device, one needs to control the motion of the propelling
machinery. For example one may need to fix its position relative to
the droplet.  One method of control is to introduce additional
steering forces acting on the device, which have to be adjusted
appropriately. As a simple example, consider a variant of the 3-point
biflagellate model, which was introduced in sect.\ref{sec:dipole}.
Here the device consists of a spherical particle of radius $a < d$
centered at $z_0\mb{e}_z$ and flagella (with negligible hyrodynamic
resistance) exerting point forces $\pm F\mb{e}_z$ at
$\mb{r}^{\pm}=(z_0\mb d)\mb{e}_z=r^\pm\mb{e}_z$, which add up to zero.
An additional force $\mb{F}^{ext}=F^{ext}\mb{e}_z$ is acting on the
sphere to guide the device (see Fig.\ref{Fig2}).

\begin{figure}
\includegraphics[width=0.45\textwidth]{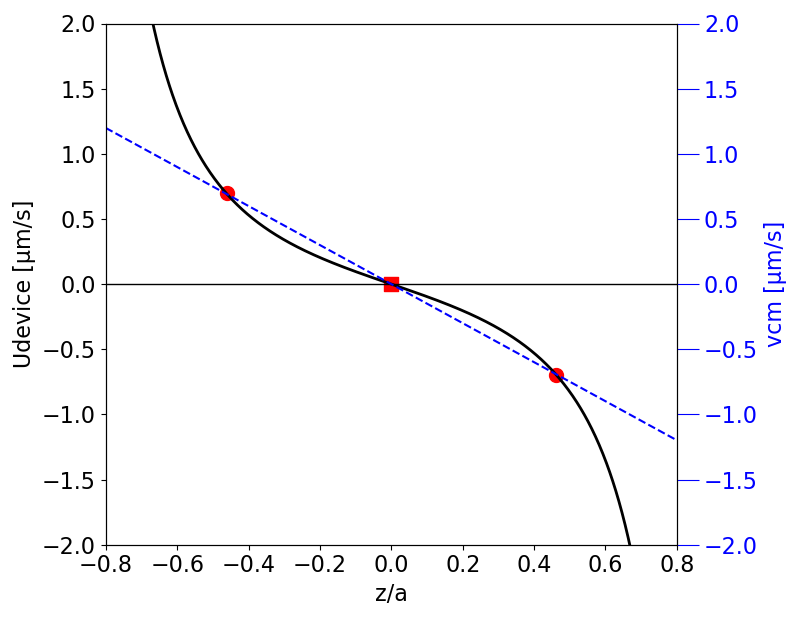}
\includegraphics[width=0.45\textwidth]{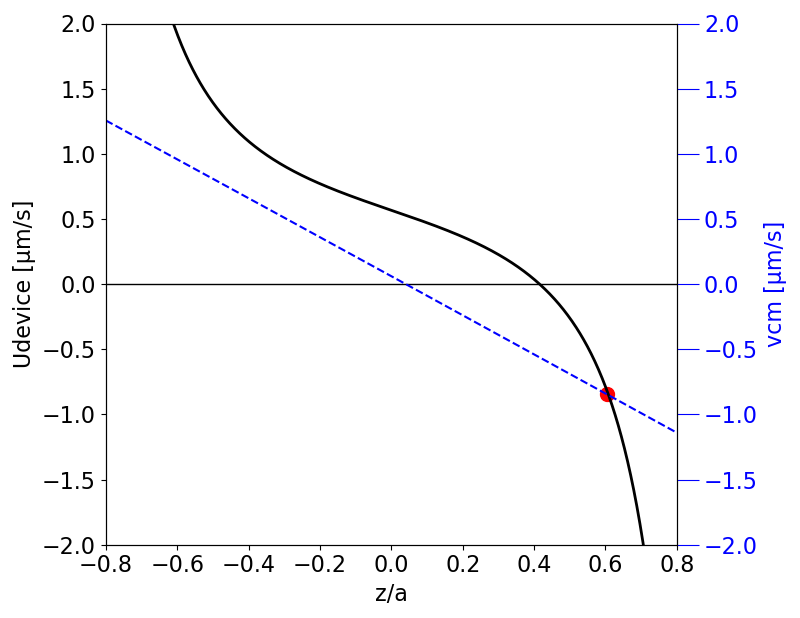}
\caption{\label{fig:guidedBiflagellate}  Device velocity $\bm{U}_{device}$ (solid line) and droplet velocity (dashed line) generated by the device described in the main text with point forces $F$ of 1 pN  at a distance $d=2\,\mu m$ from a bead with radius $1\, \mu m$, in a droplet of radius $50\, \mu m$.  (a) Without steering force $F_{ext}=0$. Stable fixpoints are marked by circles, unstable by square.  (b) With $F_{ext}=0.001\, pN$. The viscosity contrast is $\lambda=5$ and contributions up to $l=50$ are taken into account. No visible changes appear if contributions up to $l=100$ are added.} 
\end{figure}

In the absence of the sphere the point forces create a flow $\tilde{\mb{v}}$, which consists of  a sum of Stokeslets and the corresponding response flows $\mb{v}^{(\pm)}(\mb{r}|\mb{r}^{\pm})$ due to the droplet interface,  i.e.
\begin{align}
  \tilde{\mb{v}}(\mb{r})&= \Big(\mb{G}(\mb{r}-\mb{r}^+)-\mb{G}(\mb{r}-\mb{r}^-)\Big)\cdot\mb{F}\nonumber\\
  &+ \mb{v}^{(+)}(\mb{r}|\mb{r}^{+})  +  \mb {v}^{(-)}(\mb{r}|\mb{r}^{-}). 
\end{align}
Here $\mb{G}$ denotes the Oseen tensor. At $\mb{r}=\mb{r}_0$ the first term on the right hand side vanishes, because $\mb{G}$ is an even function of its argument. If we now add the spherical particle to this flow, its velocity  becomes 
\begin{align}
  \mb{U}_{device}&= \mu_0\mb{F}^{ext}
                   + \left(1 + \frac{a^2}{6} \nabla^2\right)\nonumber\\
                 &\Big(\mb{v}^{(+)}(\mb{r}_0|\mb{r}^{+})+\mb{v}^{(-)}
                   (\mb{r}_0|\mb {r}^{-}) + \mb{v}^{(a)}(\mb{r}_0|\mb{r}_0)\Big).
\label{eq:Udev}
\end{align}
This velocity contains the response flow $\bm{v}^{(a)}$ due to the particle, which  we replace by the response flow of the point  force $\bm{F}^{ext}$ as explained in the previous subsection. To leading order in $a$ the device velocity becomes
\begin{align}
  \mb{U}_{device}&= \mu_0\mb{F}^{ext}
                   + \nonumber\\
   &\mb{v}^{(+)}(\mb{r}_0|\mb{r}^{+})+\mb{v}^{(-)}(\mb{r}_0|\mb{r}^{-}) + \mb{v}^{(a)}(\mb{r}_0|\mb{r}_0).
\label{eq:Udevice}
\end{align}
The steering force can now be used to control the velocity of the device relative to the droplet velocity; the latter is explicitly given by:
\begin{equation}
\mb{v}_{cm}=\frac{1}{4\pi\eta^+}\frac{1}{3\lambda +2}\left[(2\lambda +3 -z_0^2)F^{ext}\mb{e}_z -4dz_0F\mb{e}_z\right].
\end{equation}
The relative velocity $\mb{U}_{device}-\mb{v}_{cm}$ is shown in Fig.(\ref{fig:guidedBiflagellate}). 
If $F_{ext}=0$ (see figure(\ref{fig:guidedBiflagellate} a) there are two off-center stable fixpoints of the one-dimensional motion of the 
biflagellate, whereas the center of the sphere is an unstable fixpoint. With a tiny external force  one can 
select a stable fixpoint (see figure(\ref{fig:guidedBiflagellate} b), which makes it possible to switch between forward and backward motion along the z-axis. 
With increasing $F_{ext}$ the device approaches the interface and the number of l-components needed for a correct description increases.

It is straightforward to generalise these results to more complex
arrays of point forces, including asymmetric configurations and
rotations of the droplet.

\section{Conclusions and outlook}
\label{conclusions}

Point forces in the interior of a fluid droplet generate flow and
thereby are able to propel the droplet. By constructing the Green
function for the boundary conditions of a fluid droplet, we were able
to compute the linear and rotational velocities of the droplet as
a function of general point force configurations.

Starting from a single point force, we showed that the translational
mobility of the droplet is anisotropic:
$v_{cm}^i=\mu_{t}^{ij}(\mb{r}_0) F^j $. The rotational motion of the
droplet is determined by the torque of the device with respect to the
center of the droplet, i.e. $\mb{r}_0 \times \mb{F}$, and the
rotational mobility is isotropic but potentially nonmonotonic as a
function of the position of the device. Considering stepwise more
complex configurations, we next discussed force dipoles, which in
general give rise to both, translational and rotational motion of the
droplet. Force dipoles include as special cases stresslets as a simple
model for an autonomous device as well as rotlets as an approximation
for externally driven, helical swimmers.  A more refined model of an
autonomous device includes three point forces, such that both, the
total force and torque vanish. The translational velocity of the
droplet is independent of the position of the device, if the total
force dipole moment vanishes.

We have also considered general force distributions and have shown
that the propulsion velocities of the droplet depend only on a few
multipole moments of the force. The translational velocity is
determined by the total force, dipole and quadrupole moment. The
rotational velocity requires in addition a special octupole
moment. Therefore different force distributions give rise to the same
propulsion of the droplet, provided these low order moments are the
same.

For many application, such as medical micro-robotics, it is desirable
to control the trajectory of the forcing device, such that a prescribed
trajectory of the droplet results. Approximating the device by a small spherical bead,  
allows us to use Stokes law supplemented by Fax\'en's law to derive an
approximate relation between the velocity of the device and the
applied forces. To illustrate our approach, we consider two
illustrative examples: a single bead dragged by an external force and
a biflagellate swimmer modeled as a stresslet.
The latter allows for a steady comoving state of droplet and device.
Depending on the position of the device within the droplet, we find
two stable fix points for the relative velocity and finite porpulsion.
The two stable fixpoints are separated by an unstable one, where the
device is located at the center of the droplet and no propulsion
occurs.  If the biflagellate swimmer is additionally controlled by a
tiny external steering force, one of the fixpoints is selected, allowing for a switch between forward and backward motion.

Alternatively one might consider a squirmer
in a droplet, possibly driven by an external force in addition to the autonomous activity. Thereby the trajectory of the squirmer as well as of the droplet could be controled, generalisng the work of reference \cite{Reigh2017}.
Work along these lines is in progress.

 \appendix
 \section{The Oseen tensor in vector spherical harmonics}
 \label{app:greensfct}
 
 We want to find the solution of the Stokes equation
 \begin{align}
   \label{eq:appendixStokes}
 \eta \nabla^2 \mb{v}(\mb{r} \, |\, 
 \mb{r}_0) - \nabla p + \mb{f}(\mb{r}-\mb{r}_0)&= 0 \\
 \nabla \cdot \mb{v} &= 0\nonumber
  \end{align}
 inside and outside of a droplet with appropriate boundary conditions at $r=1$ and with a point force inhomogeneity,  $\mb{f}=\mb{F}_0 \delta(\mb{r}-\mb{r}_0)$ at $\mb{r}_0$. 
 The Green's tensor with cartesian components $\mathcal{G}_{ij}$ is related to this solution via
 \begin{equation}
   \label{Green_function}
   v_i(\mb{r} \, |\, \mb{r}_0)=\sum_j \mathcal{G}_{i,j}(\mb{r} \, |\, \mb{r}_0)
   F_{j}
 \end{equation}
To simplify the notation we suppress the  $\mb{r}_0$ dependence  and write $\mb{v}(\mb{r})$,  unless stated otherwise.

Our general strategy is to construct a special solution of the
inhomogeneous equation and then add a homogeneous solution to match
the boundary conditions. 
All fields are expanded in VSH as given in
Eqs.(\ref{force_expansion},\ref{flow_expansion},\ref{pressure_expansion}). Inserting
these expansions into Eq.(\ref{eq:appendixStokes}), the Stokes
equation is converted into a system of ordinary differential equations
for the $v_{lms}$ and $p_{lm}$
modes 
\begin{widetext}
\begin{align}
  \eta \begin{pmatrix}
 \left ( \frac{\text{d}^2}{\text{d}r^2}+\frac{2}{r}\frac{\text{d}}{\text{d}r} \right )-\frac{(2+l(l+1))}{r^2} & \frac{2l(l+1)}{r^2} & 0 \\
 \frac{2}{r^2} & \left ( \frac{\text{d}^2}{\text{d}r^2}+\frac{2}{r}\frac{\text{d}}{\text{d}r} \right ) - \frac{l(l+1)}{r^2} & 0 \\
 0 & 0 & \left ( \frac{\text{d}^2}{\text{d}r^2}+\frac{2}{r}\frac{\text{d}}{\text{d}r} \right ) -\frac{l(l+1)}{r^2}
 \end{pmatrix}
 \begin{pmatrix}
 v_{lm0} \\
 v_{lm1} \\
 v_{lm2}
 \end{pmatrix}
 =\begin{pmatrix} 
 \frac{\text{d} p_{lm}}{\text{d}r} \\
 \frac{p_{lm}}{r} \\
 0
 \end{pmatrix}
 - \frac{\delta (r-r_0)}{r_0^2} 
 \begin{pmatrix}
 f_{lm0}  \\
 f_{lm1}  \\
 f_{lm2}
 \end{pmatrix}
 \label{eq:modematrix}
\end{align}
\end{widetext}

The pressure has to be determined from $\nabla^2 p=\nabla\cdot\mb{f}$, which results from the incompressibility. 
 Note that in Eq.(\ref{eq:modematrix}) the $s=0$ and the $s=1$ modes are still coupled, but we can use the incompressibility condition  $\nabla\cdot\mb{v}=0$, which becomes
\begin{align}
\label{eq:vlm1}
 \frac{\text{d}v_{lm0}}{\text{d}r}+\frac{2}{r}v_{lm0}-\frac{l(l+1)}{r}v_{lm1}=0,
 \end{align}
 to express the $v_{lm1}$ via the $v_{lm0}$ and get two decoupled equations for each $lms$-mode. 

 \begin{align}
   \eta \big( \frac{\text{d}^2 v_{lm0}}{\text{d}r^2}+&\frac{4}{r}\frac{\text{d} v_{lm0}}{\text{d}r}+\frac{2-l(l+1)}{r^2}v_{lm0} \big) \nonumber\\
   =& \frac{\text{d}p_{lm}}{\text{d}r}-\frac{f_{lm0}}{r_0^2} \delta(r-r_0)
      \label{eq:vlm0}
   \\
 \eta \big( \frac{\text{d}^2 v_{lm2}}{\text{d}r^2}+&\frac{2}{r}\frac{\text{d} v_{lm2}}{\text{d}r}-\frac{l(l+1)}{r^2}v_{lm2} \big)=-\frac{f_{lm2}}{r_0^2} \delta(r-r_0)\label{eq:vlm2}
 \end{align}
 
 Let us briefly recall the homogeneous case: all $f_{lms} = 0$.  Then,
 the Laplace equation for $p$ is solved by $p \propto r^{\ell}$ inside
 the sphere and by $p^+ \propto r^{-\ell-1}$ on the outside. For the
 quantities of the outside fluid we use a superscript $+$, i.e. the
 flow is denoted by $\mb{v}^+$, the pressure by $p^+$ and the
 viscosity by $\eta^+$.  Given the simple power laws for the pressure,
 the flow can easily be obtained, both in the ambient fluid,
  \begin{align}
 v^+_{lm0}(r)&=a^+_{lm} r ^{- \ell} +b_{lm}^+  r ^{- \ell-2} \\
 v^+_{lm1}(r)&=\frac{2-\ell}{\ell (\ell+1)} a^+_{lm}  r ^{- \ell} - \frac{1}{\ell+1}b^+_{lm}  r ^{- \ell-2} \\
 v^+_{lm2}(r)&=c^+_{lm} r ^{- \ell-1},
 \end{align}
 and in the droplet's interior,
 \begin{align}
 v_{1m0}(r)&=a^-_{lm}  r ^{ \ell+1} +b_{lm}^- r ^{\ell-1}\label{vlm0}\\
   v_{lm1}(r)&=\frac{3+\ell}{\ell (\ell+1)} a^-_{lm}  r ^{ \ell+1} + \frac{1}{\ell}b^-_{lm}  r ^{\ell-1} \label{vlm1}\\
   v_{lm2}(r)&=c^-_{lm}  r^{\ell}.\label{vlm2}
 \end{align}
 The pressure modes are then given by $p_{lm}^+ (r) = \eta^+ \frac{4\ell-2}{\ell +1} a_{lm}^+  r^{-\ell -1 }$ and 
 $p_{lm} (r) = \eta \frac{4\ell+6}{\ell} a_{lm}^-  r^{\ell}$, which completes the general solution of the homogeneous equations.

 To obtain a special solution of the inhomogeneous equation, we first
 solve Eq.(\ref{eq:pressure}) for the pressure $p$, as explained in
 the main text in Eqs.(\ref{p1},\ref{p2}).
When the solution for the pressure is plugged into Eq.(\ref{eq:vlm0}), the equation for $v_{lm0}$ becomes
\begin{align}
 & \eta \big( \frac{\text{d}^2 v_{lm0}}{\text{d}r^2}+\frac{4}{r}\frac{\text{d} v_{lm0}}{\text{d}r}+\frac{2-l(l+1)}{r^2}v_{lm0} \big) \nonumber\\
& = \ell A_{lm} r^{\ell-1} \Theta(r_0-r) - (\ell+1) B_{lm} r^{-\ell -2} \Theta(r-r_0) \nonumber
\end{align}

Again, one can construct the solution by adding inner ($r<r_0$) and outer ($r>r_0$) parts. Special solutions of the inhomogeneous equation are easily identified as 
$C_{lm}^- r^{\ell+1} \Theta(r_0-r)$ for the inner and $ C_{lm}^+ r^{-\ell} \Theta(r-r_0)$ for the outer part, respectively.  To ensure the conditions 
$v_{lm0}(r\nearrow r_0)=v_{lm0}(r\searrow r_0)$ $v'_{lm0}(r\nearrow r_0)=v'_{lm0}(r\searrow r_0)$ (here the prime denotes derivative with respect to $r$), we have to add solutions of the homogeneous equation and thus end up with an ansatz of the form 
\begin{align}
 & v_{lm0}(r)= C_{lm}^- r^{\ell+1} \Theta(r_0-r) + C_{lm}^+ r^{-\ell} \Theta(r-r_0)\nonumber\\
  &+ F_{lm} r^{\ell-1} \Theta(r_0-r) + H_{lm} r^{-\ell -2} \Theta(r-r_0)
\end{align}
Note that for $r>r_0$ we added a homogeneous solution which decays as $r\to\infty$ to ensure the boundary condition selecting the Oseen solution. 
Straightforward calculation gives the following result:
\begin{align}
C_{lm}^- &= \frac{\ell}{4\ell +6} \frac{A_{lm}}{\eta^-}\nonumber \\
C_{lm}^+ &= \frac{\ell+1}{4 \ell-2} \frac{B_{lm}}{\eta^-}\nonumber \\
F_{lm} &=  \frac{2 C_{lm}^+}{2 \ell +1} r_0^{-2 \ell +1} - \frac{2\ell +3}{2 \ell+1} C_{lm}^- r_0^2 \nonumber\\
H_{lm} &=-\frac{2\ell -1}{2\ell +1} C_{lm}^+ r_0^2- \frac{2}{2\ell+1} C_{lm}^- r_0^{2\ell +3}.\nonumber
\end{align}
The constraint of incompressibility determines
$v_{lm1}$  from Eq.(\ref{eq:vlm1}). 
Finally, by matching inner and outer solutions we obtain $v_{lm2}$ from Eq.(\ref{eq:vlm2}),
\begin{align}
  \label{v_lm2}
 v_{lm2}(r)=-\frac{f_{lm2}}{\eta^- (2\ell+1)}\left (\Theta(r-r_0)\frac{r_0^{\ell}}{r^{\ell+1}}+\Theta(r_0-r)\frac{r^{\ell}}{r_0^{\ell+1}} \right ).
 \end{align}

 The flow field, which was obtained as a special solution of the inhomogeneous equation, can be used to deduce the Oseen tensor expanded in vector spherical harmonics:
 \begin{equation}
  G_{ij}(\mb{r},\mb{r}_0)=\sum_{s,s'}\sum_{lm} G_{lm}^{s,s'}(r,r_0)
   (\mb{Y}_{lm}^{(s)}(\Omega))_i  (\mb{Y}_{lm}^{*\,(s')}(\Omega_0))_j.\nonumber
\end{equation}
Inserting this expansion into Eq.(\ref{Green_function}) yields
\begin{equation}
 v_{lms}(r,\mb{r}_0)=\sum_{s'}G_{lm}^{s,s'}(r,r_0)f_{lms'}(\Omega_0)\frac{1}{A_{l}^{(s')}}, \nonumber
  \end{equation}
  where we have made the dependence on $\mb{r}_0$ explicit again.
  The chiral component can be read off directly from Eq.(\ref{v_lm2})

\begin{align}
 & \sqrt{\ell(\ell+1)}G_{lm}^{2,2}(r,r_0)=\nonumber\\
 & -\frac{1}{\eta (2\ell+1)}\left (\Theta(r-r_0)\frac{r_0^{\ell}}{r^{\ell+1}}+\Theta(r_0-r)\frac{r^{\ell}}{r_0^{\ell+1}} \right ).\nonumber
  \end{align}
The other components require the inversion of a $2\times 2$ matrix, which is not needed here.

Having constructed a special solution to the inhomogeneous equation,
one can now simply add the homogeneous flow fields to obtain the
general solution and match the boundary conditions, which is explained
in the next section.

 \section{Matching boundary conditions}\label{boundary_conditions}
 The flow at the outside is given by the homogeneous solution
 $v_{lms}^+$, while the flow on the inside is given by the superposition of the
 inhomogeneous and the homogeneous flow. The
 boundary conditions  determine
 the six, yet unknown coefficients $a_{lm}^{\pm}, b_{lm}^{\pm}$ and
 $c_{lm}^{\pm}$
 of the homogeneous flow. There are 3 linear equations emerging from the
 continuity of the flow velocity at $r=1$, and another set of 3 linear
 equations emerging from the continuity of tractions: $\mb{e}_r\cdot(\boldsymbol{\sigma}_+-\boldsymbol{\sigma}_-)=2\gamma_0\mb{e}_r$.  As the $s=2$
 mode only involves the $c_{lm}^{\pm}$, the $6\times 6$ matrix of
 coefficients splits into a $2\times 2$ block (coupling
 $c_{lm}^{\pm}$) $\mb{\hat{M}_2}$ and a $4\times 4$ block
 $\mb{\hat{M}_0}$. The inhomogeneities in these linear equations are
 due to the special inhomogeneous solution inside. To determine all the terms in the linear equations
 we have to calculate the normal stress inside and outside of the droplet. For details of the calculation see \cite{Kree_2017}.

With $\mb{Z}_0=(a_{lm}^+, b_{lm}^+, a_{lm}^-, b_{lm}^-)^t$ and  $\mb{Z}_2=(c_{lm}^+, c_{lm}^-)^t$ the boundary conditions become $\mb{\hat{M}_0}\mb{Z}_0=\mb{I}$ and
$\mb{\hat{M}_2}\mb{Z}_2=\mb{J}$ with
\begin{align}
  &\mb{\hat{M}_0}=\nonumber\\   
  &
    \begin{pmatrix}
1 & 1 & -1 & -1\\
\frac{2-l}{l(l+1)} & \frac{-1}{l+1} & -\frac{l+3}{l(l+1)} & -\frac{1}{l}\\
2\eta^+\frac{l^2+3l-1}{l+1} & 2\eta^+(l+2) & \frac{2}{l}\eta^-(l^2-l-3) & 2\eta^-(l-1)\\
2\eta^+(l^2-1) & 2\eta^+l(l+2) & -2\eta^-l(l+2) & -2\eta^-(l^2-1)
 \end{pmatrix}\nonumber
\end{align} 
and 
\begin{equation}
\mb{I}=
\begin{pmatrix}
 C^+_{lm}+H_{lm} \\
 \frac{2-l}{l(l+1)} C^+_{lm}-H_{lm}/(l+1) \\
 B_{lm}+2\eta^-lC^+_{lm}+2\eta^-(l+2)H_{lm}\\
 2\eta^-(l^2-1)C^+_{lm}+2\eta^-l(l+2)H_{lm}
 \end{pmatrix}
\end{equation} 
as well as
\begin{equation}
\mb{\hat{M}_2}
= 
\begin{pmatrix}
1 & -1\\
\eta^+(l+2) & \eta^-(l-1) 
\end{pmatrix}
\end{equation} 
and
\begin{equation}
\mb{J}
= 
-\frac{\gamma_{lm}r_0^l}{\eta^-(2l+1)}
\begin{pmatrix}
1\\
\eta^{-}(l+2) 
\end{pmatrix}
\end{equation}   
  
The solutions for the $c_{lm}^{\pm}$  can be obtained easily by hand
\begin{align}
 c_{lm}^- &= \frac{ \frac{\ell +2 }{2\ell +2} f_{lm2} \left ( \frac{\eta^+}{\eta^-}-\frac{1}{2} \right ) \frac{r_0^{\ell}}{a^{\ell+1}}}{\eta^- (\ell-1) + \eta^+ (\ell+2)} \\
 c_{lm}^+ &= \frac{ \frac{\ell +2 }{2\ell +2} f_{lm2} \left ( \frac{\eta^+}{\eta^-}-\frac{1}{2} \right ) \frac{r_0^{\ell}}{a^{\ell+1}}}{\eta^- (\ell-1) + \eta^+ (\ell+2)} - \frac{f_{lm2}}{\eta^- (2\ell+1)} \frac{r_0^{\ell}}{a^{\ell+1}} ,       
\end{align}
whereas the $a_{lm}^{\pm}, b_{lm}^{\pm}$ obtained $\mb{\hat{M}_0}^{-1}\mb{I}$  are best manipulated by symbolic machine computation or numerical evaluation, except for the special case $\ell=1$, discussed below.\\

\section{Propulsion velocities}\label{simple_cases}

The propulsion velocities require only the $\ell=1$ component of the
flow (\ref{eq:vcm},\ref{eq:omega}).
Since the linear equations, $\mb{\hat{M}_0}\mb{Z}_0=\mb{I}$, for
the coefficients $\mb{Z}_0=(a_{lm}^+, b_{lm}^+, a_{lm}^-, b_{lm}^-)^t$
are decoupled for different $\ell$, it is sufficient to consider the
case $\ell=1$, for which the matrix $\mb{\hat{M}_0}$ simplifies
considerably. The coefficients are given explicitly by
\begin{align}\label{coeff_a}
a_{1m}^+&=(I_{1m}^3-I_{1m}^4)/(3\eta^+)\\
  b_{1m}^+&=\frac{I_{1m}^4+6\eta^-(I_{1m}^1-I_{1m}^2-a_{1m}^+/2)}{3(2\eta^++3\eta^-)}\label{coeff_b}
\end{align}
If the force is parallel to $\mb{r}_0=z_0 \mb{e}_z$, only $m=0$ contributes: $f_{100}=\sqrt{\frac{3}{4\pi}}F_0$. The linear propulsion velocity is then given by:
\begin{align}\label{vcm_parallel}
  \mb{v}_{CM}=&\,\mb{e}_z\sqrt{\frac{3}{4\pi}}(a_{10}^++b_{10}^+)\nonumber\\
  =&\frac{{\mb F}}{4\pi\eta^+}
  \frac{3\eta^++2\eta^--\eta^+z_0^2}{2\eta^++3\eta^-}.
\end{align}
In this case, the rotational velocity is given by Eq.(\ref{eq:rotationOmega}).

For a general direction of the force
the following relations are used to express the expansion coefficients $f_{1ms}$ in coordinate free form:
\begin{eqnarray}
\mb{F}\cdot \mb{Y}_{10}^0 &=& \sqrt{\frac{3}{4\pi}} \mb{F}_{||}\cdot\mb{e}_z \\
\mb{F}\cdot \mb{Y}_{10}^1 &=& \sqrt{\frac{3}{4\pi}} (\mb{F}-\mb{F}_{||})\cdot\mb{e}_z \\
  \mb{F}\cdot \mb{Y}_{11}^0 &=& \sqrt{\frac{3}{8\pi}}\mb{F}_{||}\cdot(\mb{e}_x +
                                i\mb{e}_y) \\
  \mb{F}\cdot \mb{Y}_{11}^1 &=& \sqrt{\frac{3}{8\pi}}(\mb{F}-\mb{F}_{||})
                                \cdot(\mb{e}_x +i\mb{e}_y) \\
\mb{F}\cdot \mb{Y}_{10}^2 &=& \sqrt{\frac{3}{4\pi}} \frac{\mb{r}_0\times\mb{F}}{r_0}\cdot\mb{e}_z \\  
\mb{F}\cdot \mb{Y}_{11}^2 &=& \sqrt{\frac{3}{8\pi}}\frac{\mb{r}_0\times\mb{F}}{r_0}\cdot(\mb{e}_x +i\mb{e}_y) 
\end{eqnarray}  
where
 $\mb{F}_{||}=(\mb{F}\cdot\mb{r}_0)\mb{r}_0/r_0^2$.

\bibliography{MicroswimmerNew}
\end{document}